\documentclass{llncs}

\usepackage[utf8]{inputenc}
\usepackage[T1]{fontenc}
\usepackage{csquotes}
\usepackage{hyphenat}

\usepackage{amssymb}
\usepackage{amsmath}
\usepackage{array}
\usepackage{hhline, multirow, enumitem}
\usepackage{graphicx}
\usepackage{blindtext}
\usepackage{listings}

\graphicspath{ {images/} }

\begin{document}

\title{An Exploration of H-1B Visa Applications in the United States}

\author{Habeeb Hooshmand, Joseph Martinsen, Jonathan Arauco, Alishah Dholasaniya, Bhavik Bhatt}
\institute{Capital One Tech Intern Program\\~\\
\email{ \{habeeb.hooshmand, joseph.martinsen, jonathan.arauco, alishah.dholasaniya, bhavik.bhatt\}@capitalone.com }}

\maketitle

\begin{abstract}
The H-1B visa program is a very important tool for US-based businesses and educational institutes to recruit foreign talent. While the ultimate decision to certify an application lies with the United States Department of Labor, there are signals that can be used to determine whether an application is likely to be certified or denied. In this paper we first perform a data-driven exploratory analysis. We then leverage the features to train several classifiers and compare their performance. Finally, we discuss the implications of this work and future work that can be done in this area. 
\end{abstract}

\section{Introduction}

H-1B visas are a very important route into the United States for foreign workers. The H-1B visa program allows United States employers to employ foreign workers in specialized roles~\cite{docquier2006international}. 

There are several steps to being certified for this visa~\cite{sgmimmigrationlawgroup_2017}. In order to apply for an H-1B visa, an applicant must first have an employer that is willing to sponsor their application. The applicant must already have the job guaranteed by the company, however the company may withdraw sponsorship at any time. The applicant must also submit the wage they are to receive as well as the location they are going to work. After this information has been submitted, the United States Department of Labor (USDL) must either certify or deny the application.

This application process raises two important questions. What makes an H1 candidate more likely to be certified? What candidates are frequently denied? Can we create a model that can predict whether or not a candidate is likely to be accepted?
To answer these questions, the remainder of this paper makes the following contributions:
\begin{enumerate}
  \item We explore a dataset of H-1B visa applications and their outcomes in order to determine the contributing factors to the application outcome.
  \item We train a set of classifiers to predict whether or not an application will be certified.
\end{enumerate}

\section{Related Work}

There is a plethora of related work on the H-1B visa in the Social Sciences. Lewin et al. explain how the H-1B visa is a very effective tool in combating outsourcing~\cite{lewin2009companies}. Dreher and Poutvaara show us that -- as a result of the H-1B visa -- the United States has been able to improve it's post-secondary education system by hiring top faculty from overseas~\cite{dreher2005student}. Monica Boyd has studied how recruiting tacts have changed as a result of the changes to H-1B policies and quotas over the years~\cite{boyd2014recruiting}.

There has been very little published work on the H-1B visa in the field of Computer Science. In fact, we were unable to find any published work that applied computing principles to the H-1B visa. There is one unpublished technical report titled ``H1B Visa Prediction by Machine Learning Algorithm'' found on GitHub, but it does not seem to provide any significant findings to reference~\cite{jinglin_li}. \\

\section{Data Collection and Filtering}

\noindent\textbf{Collecting the Data} The entirety of our data is provided by the United States Department of Labor's (USDL) Office of Foreign Labor Certification (OFLC) annually.\footnote{https://www.foreignlaborcert.doleta.gov/performancedata.cfm} In total the dataset provides roughly 3.1 million instances of visa applications and their outcomes. The data provides us with the following about each application: 
\begin{itemize}
  \item Case Status -- the outcome of the application
  \item Employer Name -- the name of the employer sponsoring the applicant.
  \item Job Category -- the category of the job of the applicant.
  \item Job Title -- the official job title of the applicant.
  \item Full time position -- if the position is full time, expressed as a boolean (y/n).
  \item Wage -- the wage of the applicant before taxes.
  \item Year -- the year of the application.
  \item Work site -- the city and state the applicant will work in.
  \item Latitude and Longitude -- the work site expressed as a (latitude, longitude) location.
\end{itemize}
The data provides no personally identifiable information about the applicants. The data is provided for public use by the OFLC and USDL. \\

\pagebreak

\noindent\textbf{Filtering the Data}
The raw data from the OFLC is very impure. Many of the applications are missing information. Most importantly many of the applications provided no ``Case Status''. After removing these applications from the data, $2,809,738$ applications remained. 

\section{Data Exploratory Analysis}

We perform a series of data-driven experiments in order to better understand the data from the OFLC. In order to understand our data, we ask five questions about our data. \\

\noindent\textbf{How many applications are certified?} It is important to know what percent of applications are certified and denied. This gives us a good idea of the spread of the data and whether it is more likely that a visa is certified, or if the visa is denied. 
\begin{figure}[h]
  \centering
  \includegraphics[width=0.3\textwidth]{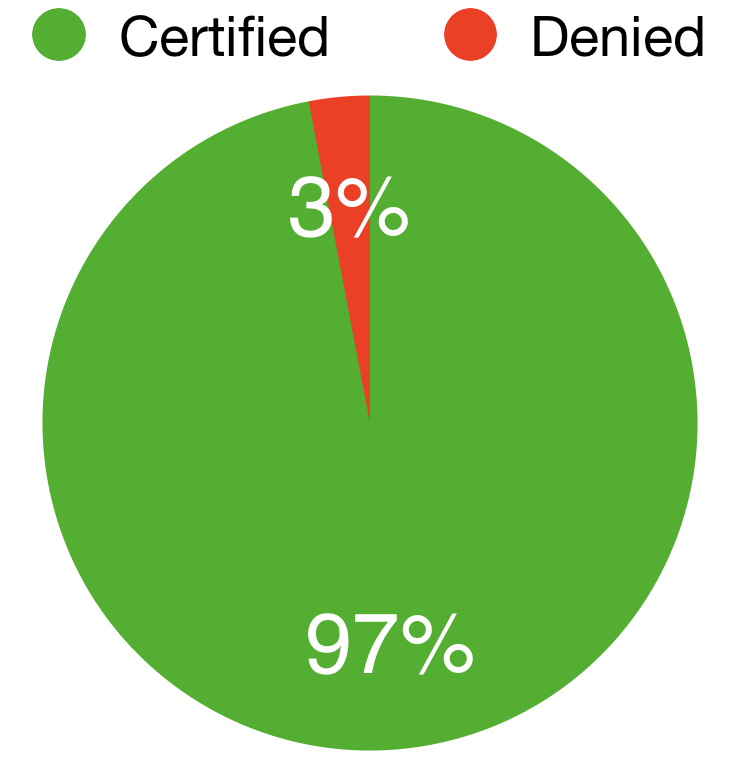}
  \caption{The certified and denied H-1B Visa applications}
  \label{fig:certified_denied_graph}
\end{figure}
Of the $2,809,738$ applications, $2,724,100$ (97\%) are certified and $85,638$ (3\%) are denied as shown in Figure~\ref{fig:certified_denied_graph}. \\

\noindent\textbf{Do wages influence applications?} To determine whether an applicant's wage influences their application's outcome, we need to first discover the spread of wages of the candidates that are certified and the spread of the wages of those who are denied. 
\begin{figure}[h]
  \centering
  \includegraphics[width=.7\textwidth]{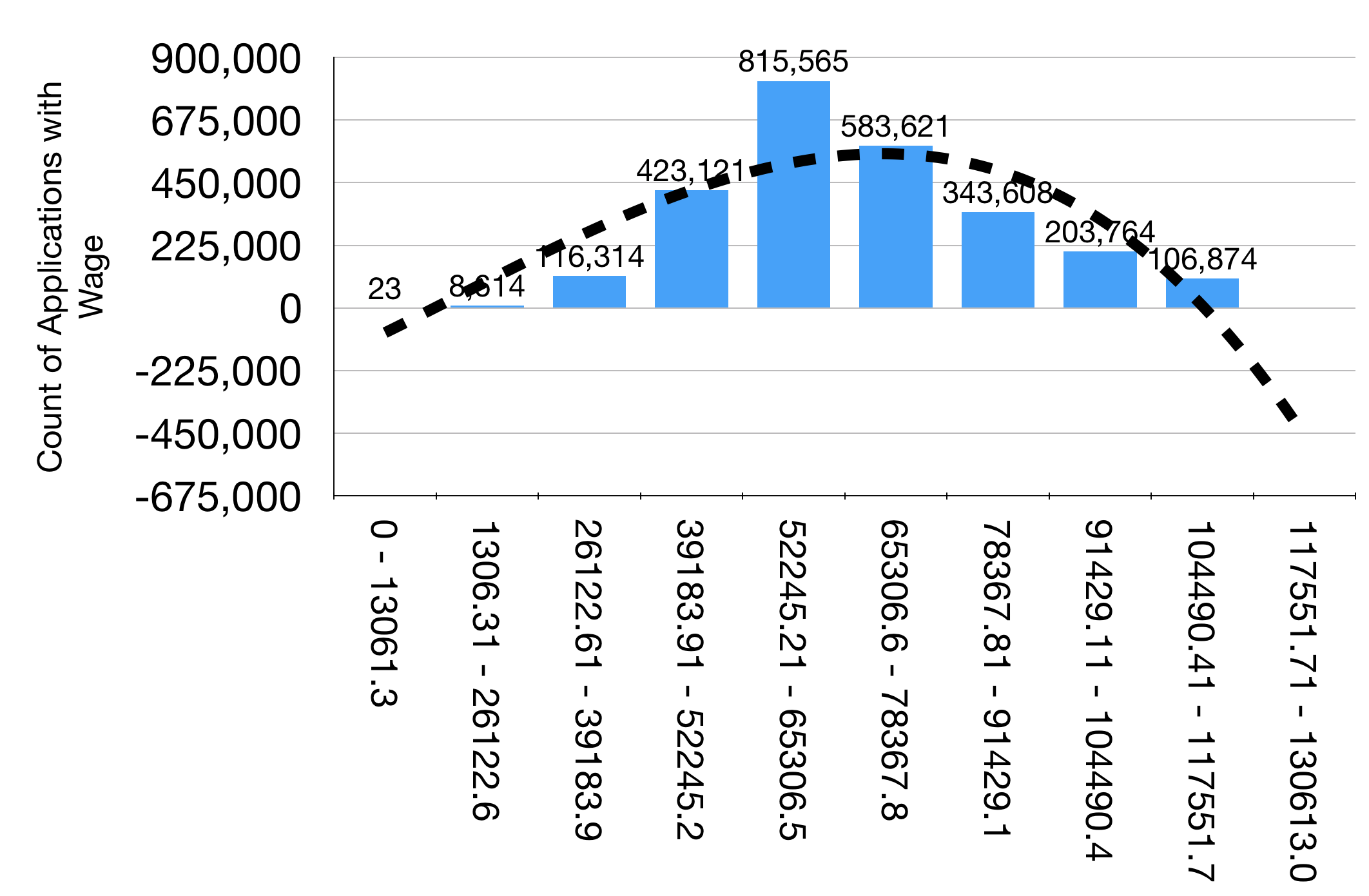}
  \caption{A histogram of certified application wages (USD)}
  \label{fig:cert_wage_hist}
  \includegraphics[width=.7\textwidth]{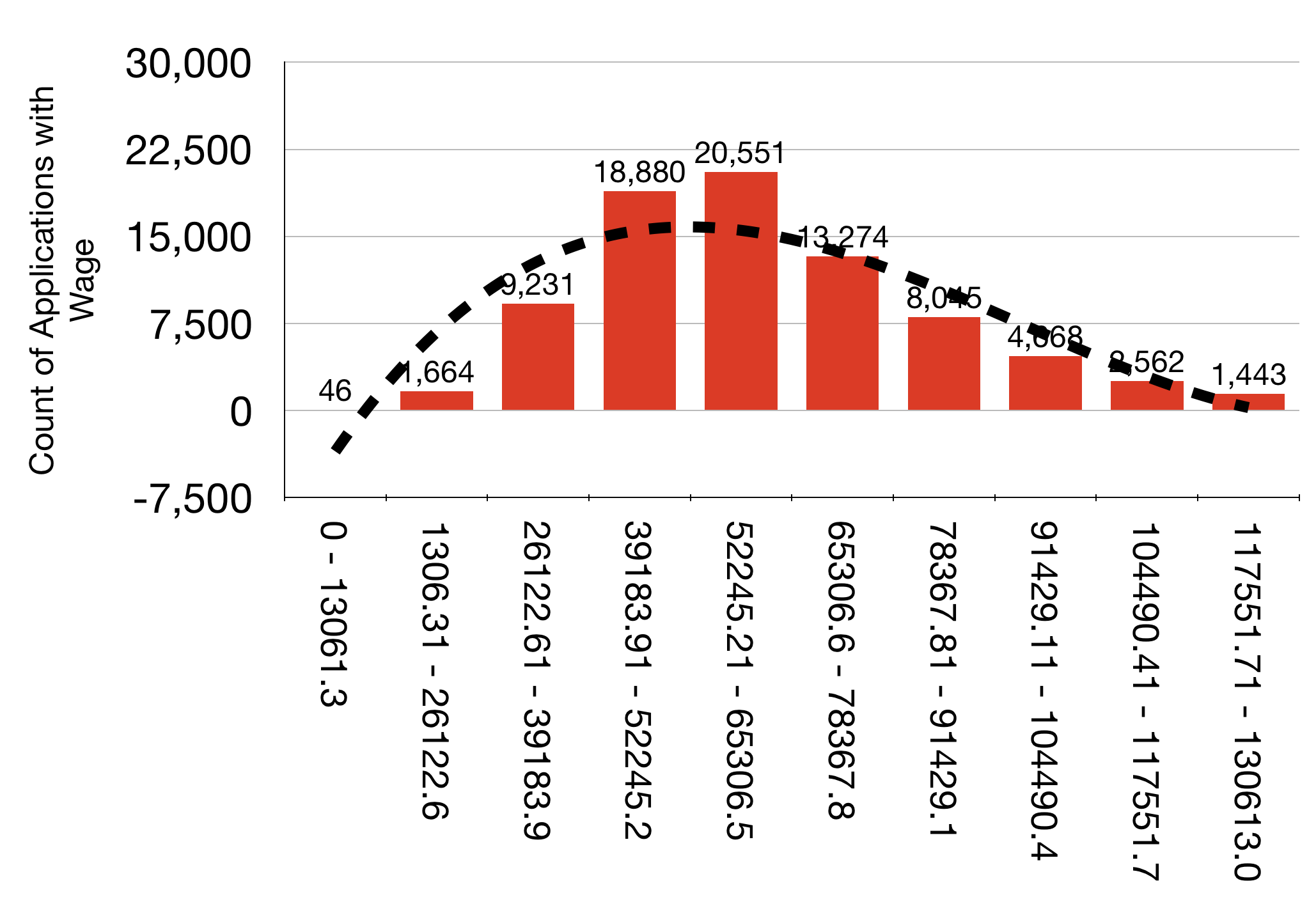}
  \caption{A histogram of denied application wages (USD)}
  \label{fig:deny_wage_hist}
\end{figure}
As shown by Figure~\ref{fig:cert_wage_hist} and Figure~\ref{fig:deny_wage_hist} the wages of certified applications are have a higher mean than the wages of denied applicants. This indicates that higher wages increase the chance of having the application approved. Meaning the wage of an applicant is a discriminative feature. \\

\begin{figure}[h]
  \centering
  \includegraphics[width=0.7\textwidth]{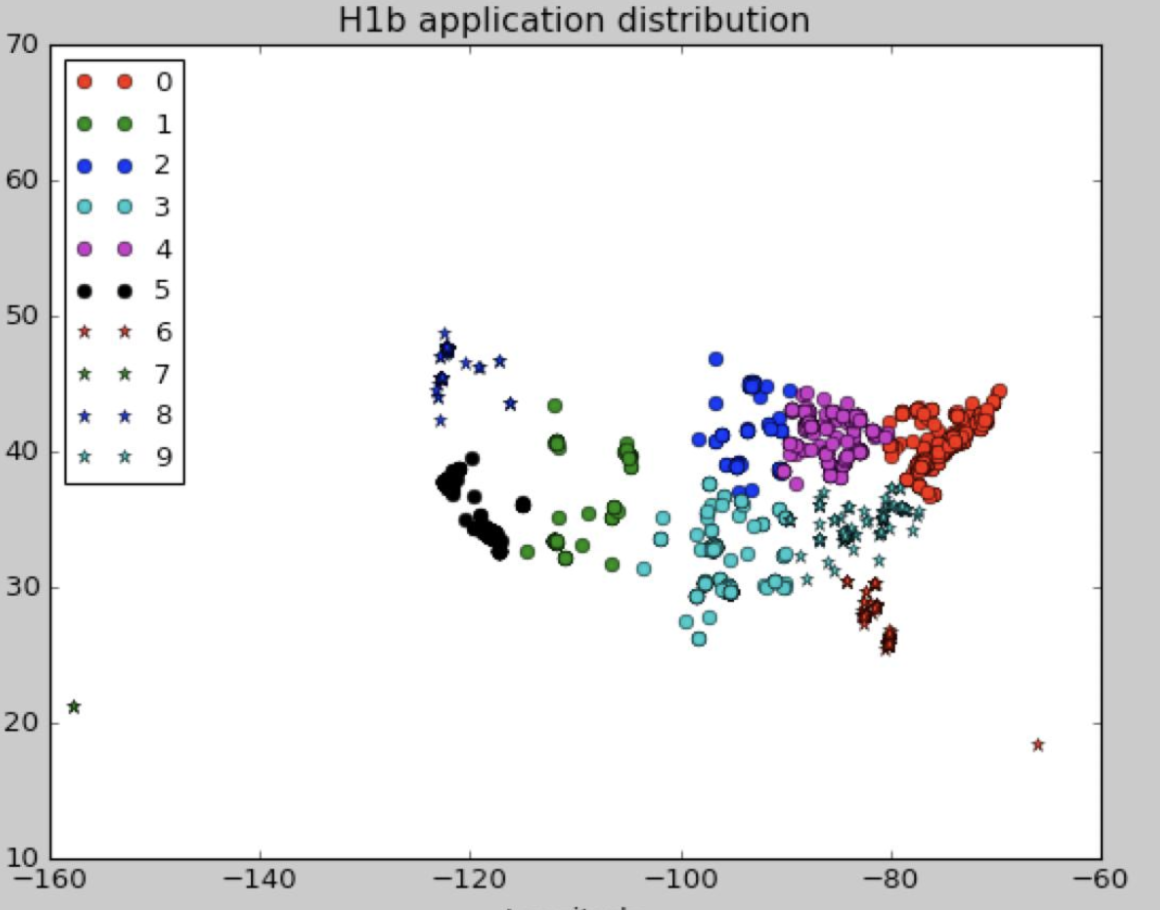}
  \caption{Locations where applicants seek work.}
  \label{fig:heatmap_apps}
\end{figure}

\begin{figure}[h]
  \centering
  \includegraphics[width=\textwidth]{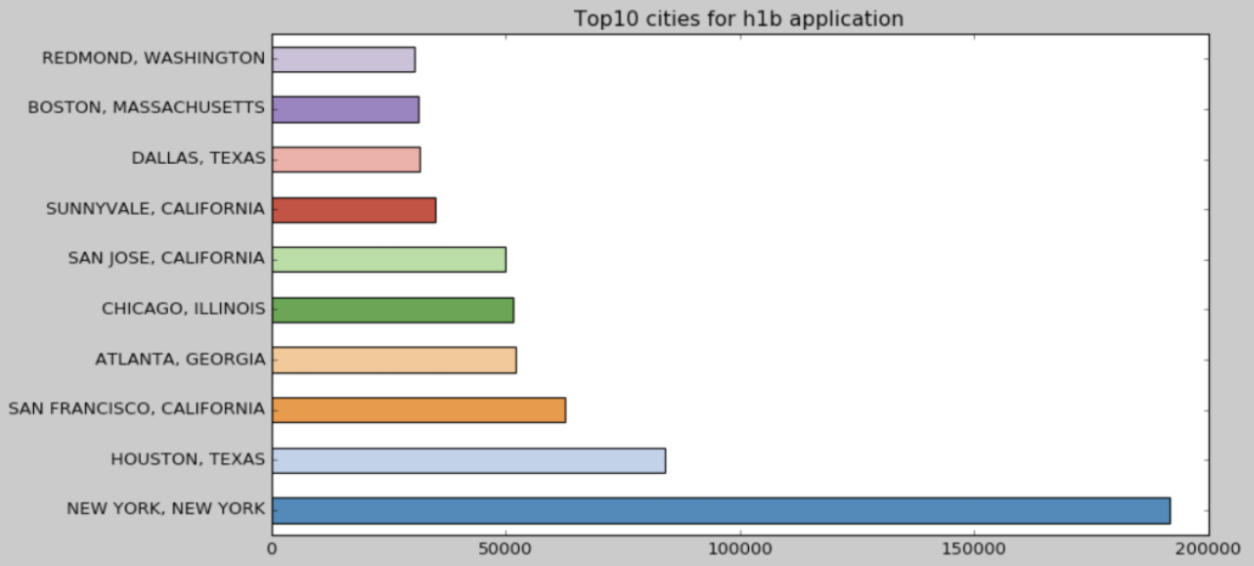}
  \caption{Locations where applicants seek work.}
  \label{fig:top_cities}
\end{figure}

\noindent\textbf{Does the work location influence an application's outcome?} If location can increase the chance of certification, many applicants could increase the likelihood of certification by seeking work in a certain region. In order to understand the influence of location on certification, we use a previously created geographic representation (Figure~\ref{fig:heatmap_apps} of all of the the H-1B visa applications that have been submitted~\cite{jinglin_li}. 
This map shows that there are locations that have employers that prefer to sponsor H-1B visa applicants. We also can see the top cities where applicants plan to work in Figure~\ref{fig:top_cities}~\cite{jinglin_li}. 
Location is clearly a discriminative feature for applicants, especially due to the fact that there are cities in which certified applicants are more likely to work.

\noindent\textbf{Does a full time position help?} In order to understand whether or not a full time job will affect the acceptance or denial of a visa application, we need to compute the difference between the applicants that have full time jobs and the applicants that do not have full time jobs as shown in Table~\ref{table:full_time}. The data indicates that a full time job is paramount to having a visa application certified. This is true to the point where only full time  workers' applications are certified in our dataset. This result may indicate missing data, but that is unlikely considering the source of the data as well as the nature of the H-1B visa~\cite{boyd2014recruiting}. This indicates that the full-time status of an applicant is a discriminative feature.\\
\begin{table} 
  \centering
  \caption{Certification and Full time jobs}
  \label{table:full_time}
  \begin{tabular}{ c | c | c }
          			& Certified 	& Denied \\
    \hline 
    Full time 		& 2,724,100 	& 0 \\ 
    \hline
    Not full time 	& 0 			& 85,638 \\
  \end{tabular} 
\end{table}

\section{Classification Task} 

\noindent\textbf{Problem Formulation} The observations from the exploratory analysis indicate that there are some features that are indicative of whether or not an H-1B visa application will be accepted. These features are:
\begin{itemize}
  \item Full time position
  \item Wage
  \item Work site
\end{itemize} 

\noindent\textbf{Problem Statement}
We have determined that these are the most discriminative features of each application. Given an application for an H-1B visa $a$, our goal is to identify whether $a$ is likely to be accepted via a binary classifier $c : a \rightarrow \{likely-certified, unlikely-certified\}$. \\

\noindent\textbf{Results} Table~\ref{table:result_data} shows the results (95\% confidence interval (CI)) of two metrics over four algorithms and five feature sets. Horizontally, both the F1-score and the AUC results are promising. While Random Forest outperforms both of the other classifiers' F1 scores, the Multilayer Perceptron Classifier performs best when considering the Area Under Curve (AUC) metric.

\begin{table}[h]
  \centering
  \caption{Classification results over two metrics and three algorithms}
  \label{table:result_data}
  \scalebox{0.75}{
		\begin{tabular}{l | c | c | c | c | c | c | c | c } 
			\hline \hline
			\multirow{2}{*}{\textbf{Feature Set}} & \multicolumn{3}{c|}{\textbf{F1 (mean)}} & \multicolumn{3}{c}{\textbf{AUC (mean)}} \\
			\hhline{~--------}
			& \textbf{KNN} & \textbf{RF} & \textbf{MLP} & \textbf{KNN} &  \textbf{RF} & \textbf{MLP}  \\
			\hline \hline
        \textbf{Combined Feature Set} & 0.9550 & \textbf{0.9848} & 0.9695 
        & 0.9768 & 0.9702 & \textbf{0.9845} \\
			\hline \hline
  \end{tabular}
  }
\end{table}

The highest AUC scores we have achieved is 0.9845, when we apply the neural network model and consider all three features at the same time. These promising results show that our classifier can leverage information from applications for H-1B visas in order to determine if an application will be accepted.

\section{Conclusion \& Future Work}
The H-1B visa program is a very important tool for US-based businesses and educational institutes to recruit foreign talent. While the ultimate decision to certify an application lies with the United States Department of Labor, there are signals that can be used to determine whether an application is likely to be certified or denied. In order to discover and use these signals, this paper provides the following contributions:
\begin{itemize}
  \item We explore a dataset of H-1B visa applications and their outcomes in order to determine the contributing factors to the application outcome.
  \item We train a set of classifiers to predict whether or not an application will be certified.
\end{itemize}

In order to accomplish our goals we first perform a data-driven exploratory analysis. Through this analysis we discover several discriminative features that we can confidently use to classify whether or not an application for an H-1B visa will be certified. We then leverage the features to train several classifiers and compare their performance. Our results indicate that these features can be reliably used to predict whether or not a visa application will be certified by the USDL.

Future work in this area would combine the existing data with data that can provide more insight into the decision made by the USDL. For instance, utilizing information about the employers that sponsor the applicants. Or, the industry field that the applicant is seeking a job in. While the topic of visas has yet to be explored in depth through the lens of Computing, this paper can provide a foundation for future work that can automate the process of visa certification.

\bibliography{bibtex}{}
\bibliographystyle{plain}

\end{document}